\pdfoutput=1

\documentclass[11pt]{article} 

\usepackage[utf8]{inputenc} 
\usepackage{geometry} 
\usepackage{graphicx} 
\usepackage{booktabs} 
\usepackage{array} 
\usepackage{paralist} 
\usepackage{verbatim} 
\usepackage{subfig} 
\usepackage{fancyhdr} 
\usepackage{sectsty}
\usepackage[nottoc,notlof,notlot]{tocbibind} 
\usepackage[titles,subfigure]{tocloft} 

\usepackage{amsmath}
\usepackage{amssymb}
\usepackage{amsthm}

\allsectionsfont{\sffamily\mdseries\upshape} 

\title{Three issues impeding communication of statistical methodology for incomplete data}
\author{JC Galati$^1$}
\date{%
    \today \\
    \mbox{}\\%
    $^1$\small{Department of Mathematics and Statistics, La Trobe University, Melbourne, VIC 3083}
}

\begin{document}
\maketitle

\begin{abstract}
   We identify three issues permeating the literature on statistical methodology for incomplete data written for
   non-specialist statisticians and other investigators. The first is a mathematical defect in the notation $Y_{obs}$,
   $Y_{mis}$ used to partition the data into observed and missing components. The second are issues concerning
   the notation `$P(R|\,Y_{obs}, Y_{mis}))=P(R|\,Y_{obs})$' used for communicating the definition of missing at
   random (MAR). And the third is the framing of ignorability by emulating complete-data methods exactly, rather
   than treating the question of ignorability on its own merits. These issues have been present in the literature for a
   long time, and have simple remedies. The purpose of this paper is to raise awareness of these issues, and to explain
   how they can be remedied.

   \vspace*{3mm}
\end{abstract}

\section{Introduction} \label{Sect:Introduction}
Missing data are a common problem in epidemiology, medicine and other fields of empirical research, and appropriate
use of statistical methods to handle the incomplete data is important. It is similarly important for users of these methods
to have a clear understanding of the concepts upon which these methods are based.

We identify and explain several shortcomings in the way these methods are communicated to non-specialist statisticians
and other investigators. Over time, these shortcomings have made their way into what now seems to be standard
practice. Their presence severely impedes dissemination of core conceptual information. The aim of this paper is to
identify and explain these issues in order to raise awareness of them, both for readers of the current literature, and for
future authors, and to communicate simple remedies for them.

We trust that this work will be received in the spirit in which it is given, namely to remove potential barriers to
understanding for users of these important statistical methods, and thereby to contribute to an increase in the
efficiency with which users acquire an understanding of the affected concepts, and ultimately to contribute to
improvement in the quality of analyses performed in practice.

\section{The core framework for modelling incomplete data} \label{Sect:Framework}
The modern framework for handling incomplete data has three components: (i) a data generation process for the
data encompassing both observed and unobserved data, and often represented by a random vector for data on all
units jointly, say~$Y$, which we will call the \textbf{data process}, (ii) a corresponding process that conceals data
values from the investigator, represented by a binary random vector, say~$R$, of the same dimension as~$Y$, and
(iii) a model of joint probability distributions for the pair of random vectors $(Y, R)$. We call the conditional
distribution of $R$ given $Y$ the \textbf{missingness process}.  This framework was introduce by Rubin\;(1976).
When the data comprise independent and indentically distributed (IID) observations on $n$ units, then we have
$Y=(Y_1, Y_2, \ldots, Y_n)$ and $R=(R_1, R_2, \ldots, R_n)$ with $n$ identically distributed pairs of random vectors
$(Y_i, R_i)$ for $i=1, 2, \ldots, n$.

\vspace*{2mm}
\textbf{Note.} It is conventional in the statistics literature to use $n\times 1$ column matrices for vector quantities.
This convention dates back at least to Halperin\;et.\,al.\,(1965), and stems from the abundance of matrix calculations
required in the theory of statistical methods. Since we have no need to perform matrix calculations, we will dispense
with this convention, and the need for transposes in setting $Y=(Y_1^T, Y_2^T, \ldots Y_n^T)^T$.

\vspace*{2mm}
\textbf{Note.} It seems common in the literature on incomplete data statistical methodology to use upper case letters,
$Y$ for example, to denote both random variables (which are functions) and their realisations (which are numbers),
and likewise for random vectors and their realisations. This seems to be a hangover from the sampling survey
methodology literature out of which much of the core ideas grew (Rubin (1987), for example). We follow the
recommendation in Halperin\;et.\,al.\,(1965) and distinguish these by reserving uppercase letters for random vectors,
and corresponding lowercase letters for realisations. For example, a dataset of observed and unobserved values
realised from $Y$ is denoted~$\mathbf{y}$.

\vspace*{2mm}
Rubin (1976) gave conditions under which the conditional densities of $R$ given $Y$ could be discarded from a
model $\mathcal{M}_g \,=\, \{\, f_\theta(\mathbf{y})\,g_\psi(\mathbf{r} |\, \mathbf{y}) : (\theta, \psi)\in\Delta \,\}$
of joint densities for~$(Y, R)$, and identical inferences could instead be derived from the simpler model
$\mathcal{M}_s \,=\, \{\, f_\theta(\mathbf{y}) : \theta\in\Theta \,\}$ for $Y$ alone. We will call the model
$\mathcal{M}_g$ the \textbf{full model}, the model $\mathcal{M}_s$ the \textbf{data model}, and the set of
conditional densities $\{ g_{\psi}(\mathbf{r}|\,\mathbf{y})\,\}$ the \textbf{missingness model}. We call each density
in the missingness model a \textbf{missingness mechanism}.

\section{The notation $Y_{obs}$, $Y_{mis}$ and its variants (Issue~1)} \label{Sect:TwoNotionsOfMissingness}
The notation $Y_{obs}$ and $Y_{mis}$ to denote the partition of $Y$ into observed and missing components
according to some realisation of $R$ originated in early works on statistical methods for incomplete data. This appears
to have started in Rubin (1976) with the notations $U_{1}$, $U_{0}$, which evolved into $Y_{obs}$, $Y_{nob}$ in
Rubin\;(1987) and $Y_{obs}$, $Y_{mis}$ in Little and Rubin\;(1987), Schafer\;(1997) and Little and Rubin\;(2002).
By the time the second edition of Little and Rubin's book appeared, use of $Y_{obs}$ and $Y_{mis}$ seems to have
become entrenched in the literature.

It has been suggessted that the meaning of $Y_{obs}$ and $Y_{mis}$ is clear when the realisation of $R$ giving rise
to the partition is fixed, and more care is needed when $R$ is considered to be variable
(Molenberghs\;et.\,al.\,(2015,\;chap.\,12)). However, this is not true. As will be explained below, the core problem with
the notation is that there are two different relationships between $Y$ and~$R$ which need to be expressed, but the
same notation is used for both (and this is the case whether $R$ is held fixed or allowed to vary).

The first relationship is the formal one of missingness defined by the random vector $(Y, R)$ in which the components
are partitioned into $(Y_{obs}, Y_{mis}, R)$, and the partition of the $Y$ components varies with~$R$. This
relationship expresses  the intended meaning of the binary random vector~$R$, whereby for any triple
$(Y_{obs},Y_{mis},R)$, the realisations of $Y_{obs}$ are always observable and the realisations of $Y_{mis}$ are
never observable. This is a static, irrevocable relationship. If $\mathbf{r}_1, \mathbf{r}_2, \ldots \mathbf{r}_k$ are
the different patterns of missingness realisable from $R$, then each missingness pattern gives rise to a different
partition:
\begin{equation} \label{Eq:Partitions}
   (Y_{obs},Y_{mis},\mathbf{r}_1), \;\;
   (Y_{obs}, Y_{mis}, \mathbf{r}_2), \;\;
   \ldots,\;\;
   (Y_{obs},Y_{mis},\mathbf{r}_k).
\end{equation}
To change the value of a component of $Y$ from observed to missing, or vice versa, the given triple must be
discarded and a new triple selected with a different pattern of missingness. Additionally, the different partitions
(\ref{Eq:Partitions}) piece together over all missingness patterns to give a single partition for all of~$(Y, R)$.

The second relationship arises when proving that $P(Y_{mis}|\,Y_{obs}, R)=P(Y_{mis}|\,Y_{obs})$ when the
missingness process is MAR with respect to some realised pair $(\mathbf{y}, \mathbf{r})$, and the distributions are
conditioned on $\mathbf{y}$ and~$\mathbf{r}$. Before this MAR equation can be proved, the distribution on the
right hand side must be defined. This involves a second, different relationship between $Y$ and~$R$. In this case,
the missingness pattern $\mathbf{r}$ is overlayed on top of the \textit{marginal distribution} for~$Y$, which has no
concept of `observed' or `missing.'  The variables $Y_{obs}$ and $Y_{mis}$ are those that were observed and
missing  `this time,' respectively, but the distribution of $Y_{obs}$ within the marginal distribution for $Y$ represents
a mixture of observable and unobservable values, and likewise for~$Y_{mis}$. Moreover, identifying a definite set of
variables $Y_{obs}$ of $Y=(Y_{obs},Y_{mis})$ on which to condition in the marginal distribution for $Y$ requires
holding the missingness pattern $\mathbf{r}$ fixed and allowing the marginal distribution of $Y$ to vary, in violation
of the stochastic relationship encoded in the random vector~$(Y, R)$.

Fortunately, the remedy for this problem is straightforward. Different notation is needed for the different relationships.
For reasons we do not delve into here, making explicit the dependence of the partition on the missingness pattern
$\mathbf{r}$ is needed as well. And to cater for the case of IID data, it helps to use superscripts instead of subscripts.
This gives two pairs of notation:
\begin{alignat}{2}
   &Y^{ob(\mathbf{r})},\;Y^{mi(\mathbf{r})} \qquad
      &&(\text{for the formal relationship $(Y, R)$}\,) \label{Eq:YFormal} \\
   &Y^{ot(\mathbf{r})},\;Y^{mt(\mathbf{r})} \qquad
      &&(\text{for the temporal relationship of overlaying $R$ onto~$Y$}).  \label{Eq:YTemporal}
\end{alignat} 
There is one subtlety with the MAR equation which needs explanation. The correct notation on the right hand side of
the MAR equation is $P(Y^{mi(\mathbf{r})}|\,Y^{ob(\mathbf{r})})$ and \textbf{not}
$P(Y^{mt(\mathbf{r})}|\,Y^{ot(\mathbf{r})})$. The reason for this is that the domain of the this function is not
$Y$ but $(Y, R)$, and the function on the right hand iside is the composition of
$P(Y^{mt(\mathbf{r})}|\,Y^{ot(\mathbf{r})})$ with the projection $(\mathbf{y},\mathbf{r})\mapsto\mathbf{y}$.
Nevertheless, to define this function properly, both notations are required, and this is true for other derivations in
the theory too.

There is more that can be said here. But this issue is treated in detail in Galati\;(2019a), where the notations
(\ref{Eq:YFormal}) and (\ref{Eq:YTemporal}) are carefully defined, and the two relationships are termed
\textbf{formal missingness} and \textbf{temporal missingness}, respectively.

\section{The notation $P(R|\,Y_{obs},Y_{mis})=P(R|\,Y_{obs})$ (Issue~2)} \label{Sect:PRYobs}
The notation $P(R|\,Y_{obs},Y_{mis})=P(R|\,Y_{obs})$ for the definition of MAR was introduced in Little and
Rubin\;(1987) and repeated in Schafer\;(1997). The main problem with this  notation is that it is impossible to interpret
because the function $P(R|\,Y_{obs})$ on the right hand side is undefined. A secondary problem is that, even once
this function is defined, the equality cannot be interpreted as a statement of MAR unless the reader knows
\textbf{not to} treat the two functions being compared as conditional probability distributions for~$R$.

It is not surprising that under these circumstances the definition of MAR has been misinterpreted (Mealli and
Rubin\;(2015)). The situation has not been helped by Little and Rubin\;(1987,\,p.\,90) having stated that
``\textit{the \textbf{distribution} of the missing-data mechanism does not depend on the missing values $Y_{mis}$}''.
This statement is simply incorrect. The definition in Rubin (1976) requires only that the \textit{probability} of the
\textit{observed} missingness pattern does not depend on~$Y_{mis}$. This restriction does not prevent the probability
 of other missingness patterns, and thereby the distribution of~$R$, from varying with~$Y_{mis}$. Similar incorrect
statements using the word `distribution' were repeated by Schafer\;(1997,\,p.\,11) and Little and Rubin\;(2002,\,p.\,119).

If the notation in the statement is interpreted to be referring to conditional probability distributions for~$R$, and in
order to make sense of the right hand side of the equality, one interprets missingness on this side of the equation to be
temporal (extended to all of $(Y, R)$, see Galati\;(2019a) for details) rather than formal missingness, then the notation
is precisely a statement of conditional independence between $R$ and~$Y_{mis}$. This `conditional independence'
interpretation of MAR now seems widespread, and it has been repeated as recently as 2015
(Molenberghs\;(2015,\,p.\,8), for example).

Given that more than three decades have elapsed since the publication of Little and Rubin (1987), it is almost certain
that the undefined notation `$P(R|\,Y_{obs})$' has proliferated substantially through the scientific and other
academic literature. Whether one considers this to be an issue is a subjective judgement. While one option is to
discourage use of this notation going forward, a better option is to formulate a definition for `$P(R|\,Y_{obs})$'
so that the existing statements `$P(R|\,Y_{obs},Y_{mis})=P(R|\,Y_{obs})$' can be interpreted mathematically in a
manner consistent with Rubin\;(1976). Future authors can then decide for themselves whether this method of stating
the definition is natural, or whether a more direct statement is preferable. This `patching' of the literature, together
with a detailed conceptual description of MAR is treated in detail in Galati\;(2019b). For the convenience of the reader,
we sketch the ideas of the former below.

So how can $P(R|\,Y_{obs})$ be defined to enable `$P(R|\,Y_{obs},Y_{mis})=P(R|\,Y_{obs})$' to be interpreted as
a definition of MAR? The first thing a reader must understand is not to treat either side as a conditional distributon
for~$R$. Rather, given a fixed missingness pattern~$\mathbf{r}$, one holds both sides fixed at $R=\mathbf{r}$, and
treats each side as a function of~$\mathbf{y}$. The definition of the left hand side (as a function of $\mathbf{y}$) is
clear, but one cannot say the same for the right hand side. If MAR holds, then as $Y_{obs}$ is held fixed and
$Y_{mis}$ allowed to vary, the function on the left is constant. In this case, the right hand side can be defined to be
equal to the left. However, when MAR does not hold, as $Y_{mis}$ varies, the function on the left hand side varies,
and there is a set of probabilities
\begin{equation}
   \mathcal{S} \;=\;
      \{\,P(\mathbf{r}|\,\mathbf{y}_{obs},\mathbf{y}_{mis}) \,:\,
      \mathbf{y}_{mis}\text{ varies over all possible values}\,\}
\end{equation}
containing more than one value. To detect this fact, $P(R|\,Y_{obs})$ must be a constant function of~$Y_{mis}$,
defined so that it will be equal to the single value $P(\mathbf{r}|\,\mathbf{y}_{obs},\mathbf{y}_{mis})$ when MAR
holds. One straightforward option is to take $P(R|\,Y_{obs})=\text{sup}\;\mathcal{S}$, but there are many other
options. For practical purposes, one can simply consider $P(R|\,Y_{obs})$ to be the largest of the probabilities
$P(R|\,Y_{obs},Y_{mis})$ that occur while holding $R$ and $Y_{obs}$ fixed and allowing $Y_{mis}$ to vary.

\section{Framing ignorability by emulating complete-data methods (Issue~3)} \label{Sect:Ignorability}
Rubin (1976) identified conditions under which direct likelihood inferences or Bayesian inferences about $\theta$
drawn from the data model will be identical to those drawn from the full model. Rubin also gave conditions under
which two sampling distibutions will be the same, which will not concern us here. 
 
For direct likelihood inferences, there are two ignorability conditions: (i) the parameter space of the full model is a
direct product of parameter spaces for the data model and the missingness model, $\Delta=\Theta\times\Psi$, called
\textbf{distinctness of parameters}, and (ii) each missingness mechanism in the missingness model is
\textbf{missing at random} (MAR). The definition of MAR (with respect to some realised pair
$(\mathbf{y}, \mathbf{r})$) is that the missingness mechanism
$g(\mathbf{r}|\, \mathbf{y}^{ob(\mathbf{r})}, \mathbf{y}^{mi(\mathbf{r}})$ , when considered as a function of
$\mathbf{y}$ with $\mathbf{r}$ fixed, is a constant function of $\mathbf{y}^{mi(\mathbf{r})}$.
For Bayesian inferences, a third condition is added whereby the prior distribution for $\theta$ and $\psi$ factorizes
into a prior distribution for $\theta$ and a prior distribution for~$\psi$. These conditions are now well known, and
we will refer to them as the \textbf{standard conditions} for `ignorability' of the missingness model. Galati\;(2019c)
calls the full models satisfying these conditions \textbf{ignorable models}.

In addition to the standard conditions for ignorabiltiy, Seaman\;et.\,al.\,(2013,\;p.\,266) observe that a different
interpretation of ignorability lurks in the literate, based on consideration of ignorable likelihood estimation on its
own merits, independent of any missingness model for the missingness process. Galati\;(2019c) reviews both
approaches to ignorability, developing the latter to incorporate direct likelihood inferences and consideration of
non-distinct  parameters. The main points are summarised below.

With complete data, model-based paradigms posit a model for the data vector $Y$ and evaluate the validity of the
model by checking the goodness of fit of the model to the data. Assumptions about the (unknown) distribution of
$Y$ are not of concern. The frequentist likelihood paradigm extends direct likelihood by further asserting that the
model is correctly specified, meaning that a density for the distribution of $Y$ is contained within the model. In this
way, an assertion about the distribution for $Y$ is made, but only indirectly via the model.

When there is no prior information to incorporate into the analysis, with regard to the missingness process when
the data are incomplete, any advantage that model-based paradigms might enjoy over frequentist paradigms with
complete data is essentially lost due to the impossibility of validating the model for the missingness process against
the observed data. Therefore, any rationale for choosing a model for the missingness process over asserting
directly properties of the conditional distribution $R$ given $Y$ is lost as well. Framing MAR in terms of the model
for the missingness process has two disadvantages. Firstly, the causal link between the model and the estimator is
partially severed. Specifically, changes in the missingness model have no bearing on the properties of the ignorable
likelihood estimation. While it is true that swapping one ignorable missingess model for another simply rescales the
likelihood by a constant, this requires a user of the tools to have a more detailed knowledge of the framework than
necessary. The second is that it leads to a somewhat convoluted answer to the question posed: to \textbf{not use}
a full model for the analysis, the investigator is directed \textbf{to choose} a full model (with specific properties).

The disadvantages just mentioned can be avoided by defining MAR directly as a property of the distribution for
$R$ given~$Y$ rather than in terms of some hypothetical model for this distribution. The distinctness of parameters
criterion is still required, but this is best framed as non-distinctness of parameters, because it represents a choice by
the investigator to exclude from the analysis specific pairs of distributions for the data and the missingness process,
rather than a mathematical property that is required to make ignorable likelihood estimation `work.' With these
changes, the two scenarios in which a full model is needed are (i) the conditional distribution for $R$ given $Y$ is
non-MAR, or (ii) the investigator wishes to impose a relationship $\Delta\subsetneq\Theta\times\Psi$ on the estimation
of~$\theta$. When neither of these apply, ignorable likelihood estimation is appropriate and equates to using the
(unknown) conditional distribution $R$ given $Y$ directly in the analysis.

The difference between the standard conditions and the ones just discussed can be summarised as follows: the
standard conditions require that for an investigator to not use a full model in the analysis, the investigator chooses to
use an ignorable (full) model; the latter states that the ignorable models are the full models that can be ignored because
conditions under which an investigator would choose to use one never arise. Alternatively, choosing an untestable
model is no better than making a direct untestable assertion about $R$ given~$Y$, and the former has several
disadvantages which the latter does not share.

\section{Discussion} \label{Sect:Discussion}
We have pointed out several shortcomings in the literature on statistical methods for incomplete data, and we have
explained how these can be overcome. While it is not the custom in statistics to openly criticise others' work publicly
in this way, these issues have persisted in the literature for so long that not doing so is unlikely to solve the problem.
We trust that this work will be accepted in the spirit in which it is given, namely, to pin-point issues that tripped-up the
author when coming to incomplete data methods from a non-statistical background, and to mitigate against others
falling into the same traps in future.


\end{document}